# Higher order optical resonance node detection of integrated disk microresonator

M. Ostrowski[1,3], P. Pignalosa[1,2], H. Smith[3] and Y. Yi[2,3,1*]

[1] New York University, New York City, New York
[2] City University of New York, SI/GC, New York City, New York
[3] Massachusetts Institute of Technology, Cambridge, Massachusetts
*Corresponding author: yys@alum.mit.edu

We have demonstrated higher order optical resonance node detection by using an integrated disk microresonator from through port of the coupling bus waveguide. In addition to the fundamental mode, disk resonator has higher order whispering gallery modes. The excited $2^{nd}$ order higher order mode has a node at the position where the electromagnetic energy of the fundamental mode is close to a maximum. This high resolution measurement of optical resonance mode profile has variety of applications for optical sensing and detection. The self-referencing characteristics of the two optical resonance modes have potential to achieve optical detection independent of external perturbation, such as temperature change.

Microresonators have been gaining increasing interests recently due to their unique optical properties and potential applications on telecommunications and sensing [1-4]. Optical detection and sensing, such as biosensing, chemical sensing, and gas sensing using optical methods are becoming more important as fast, non-contact, and potentially label-free techniques (without labeling of query molecules with fluorescent dyes) are desired. In recent years, many novel photonic structures and materials have been developed to make very sensitive devices. Currently the idea of using microresonators as sensors is attracting the attention of many groups, since the light trapped in the microresonator circulates many times (high Q) so that an effective enhancement of the interaction between the analyte and the input signal can be achieved [5-8]. A variety of types of optical microresonators have been investigated for this purpose, but microspheres, microrings and microdisks have received the most attention [9-11]. Since the fabrication process for integrated ring and disk resonators is compatible with standard microelectronic processes, these devices offer the potential for low cost and robust systems.

In order to achieve ultra sensitive detection of lower concentration and smaller objects (such as biological molecule, protein, bacteria, virus and nano scale particle) using integrated microresonators, the understanding of fundamental and higher order modes is important for such applications. Different from single mode ring microresonator, in which only one whispering gallery mode (WGM) is excited, for *disk* resonator, both the fundamental mode and the $2^{nd}$ order whispering gallery mode (WGM) are normally excited by coupling to the single mode bus waveguide. The different characteristics of the two WGMs may bring us new sensing and detection mechanism.

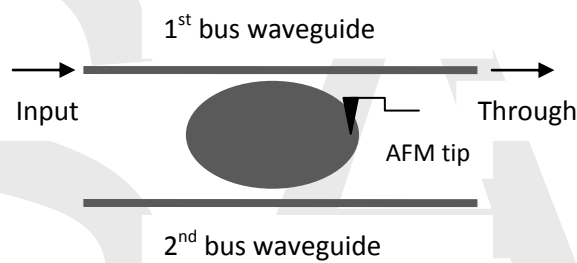

Fig. 1 The integrated disk microresonator coupled with two bus waveguides.

Here we report a different approach to detection and sensing, in which self referencing mechanism can be achieved by simultaneous detection of both fundamental and $2^{nd}$ order microdisk optical resonance modes. Additionally, we are able to measure the area around the maximum of the fundamental WGM and the node of the higher order WGM, which have overlaps in the disk. In this work, we used an on-chip disk microresonator as the example, as a variety of types of optical microresonators have been investigated (Fig. 1); we used Atomic Force Microscopy (AFM) tip to simulate the nanoparticle to realize the measurement of the two whispering gallery modes excited by the coupling bus waveguide, where the small tip can be either dielectric materials or metallic materials. A gold AFM tip was used in our experiment.

The integrated disk microresonator device is composed of a disk resonator and one or two coupled bus waveguides for light injection and signal extraction. A wavelength tunable laser was placed at the input port of the 1st bus waveguide to launch the waveguide mode; the photo detector was placed at the through port of the bus waveguide to receive the output signal after the coupling of the bus waveguide and the disk microresonator (Fig. 1). A detection of the excited whispering gallery modes event occurs when an AFM tip couples with the mode field on the disk, causing changes in the frequencies of the disk resonant modes, and coupling between the disk modes. We found the disk resonator not only possesses most of the advantages of ring resonator, which has been extensively studied in previous works, but also has the capability to make a self-referencing optical sensing and detection device due to the differential coupling of different Whispering Gallery Modes (WGM) to attached AFM tip. Because different modes have different mode profile and thus different coupling to the simulated nanoparticle, the magnitude of the induced shift depends on the specific resonant mode. Thus in cases where two or more modes are excited, relative shifts of the resonant mode wavelengths can be used in a self-referencing way.

An on-chip photonic device configuration of a microresonator with two bus waveguide has been fabricated. The thickness of Si waveguide is 220 nm. The disk resonator is about 3.6 µm in diameter with the waveguide width of 200 nm. The core material of the disk is Si with a refractive index of 3.48 at around 1.55 µm wavelength, and the bottom cladding material is $SiO_2$ with a refractive index of 1.46 [12]. We used SOI wafer for the integrated disk microresonator fabrication, and an e-beam is utilized to fabricate the 100 nm size gap between the bus waveguide and the Si disk microresonator.

The results on the measurement of the fundamental and 2nd order whispering gallery modes are illustrated in Fig. 2, shifts in the optical wavelengths corresponding to resonant whispering gallery modes of the disk are induced by proximity of a gold AFM tip. Fig. 2a is the through port signal in the wavelength range from 1.5 µm to 1.6 µm, which shows the excitation of two disk microresonator modes, the fundamental mode is at the left side with wavelength at 1549.6 nm (Q=802), the 2nd order mode is on the right side with longer wavelength at 1569 nm (Q=1180). It is interesting to observe the different modal effects of disk microresonators with and without a gold AFM tip approaching the disk. Higher Q from better fabrication control will make a more sensitive device.

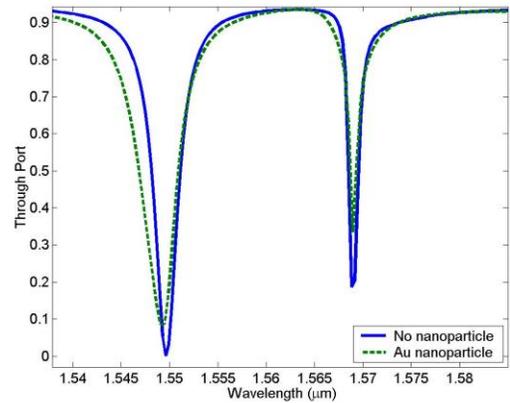

(a)

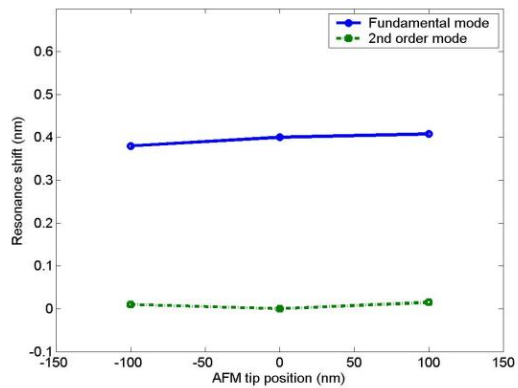

(b)

Fig. 2: The higher order optical resonance node detection (a) the fundamental mode at 1549.6 nm, the 2nd order mode at 1569 nm, the solid curve - without a gold AFM tip, the dashed curve – when the tip is close to the nodal position of the 2nd order mode. (b) The resonance wavelength shift for the two modes when the AFM tip is moved around the nodal position along the radius direction.

As illustrated in Fig. 2a, the solid curve is the through port signal without AFM tip close to the disk microresonator, while the dashed curve is the through port signal when a gold AFM tip is staged close to the surface of the disk resonator, where the tip position is at a point that is close to the maximum electromagnetic field intensity of the fundamental mode, while almost at the nodal point of the 2nd order mode. We observed about 0.4 nm resonance wavelength shift of the through port signal for the fundamental WGM mode around 1.55µm (solid curve in Fig. 2a). For the higher order WGM mode, the change induced by the attached gold AFM tip is almost zero for through port, which can be clearly seen in Fig. 2a (compared to the shift between the solid curve and the dashed curve). The differential effect for multiple WGMs indicates the advantages of the *disk* resonator – we can make it a self-referencing detection and sensing device, which can reference out frequency shifts induced by effects other than the nanoparticle interaction, e.g. changes in the refractive index of the medium surrounding the resonator, such as the fluctuation of temperature. Although we only use a gold AFM tip as an example, other nanoparticles with high refractive index and large absorption, such as Si, Al and Ag nanoparticles, should possesses similar characteristics. Fig. 2b is the results when the AFM tip is moved along the radius direction around the position in Fig. 2a, the solid line is the resonance wavelength shift from the fundamental mode, while the dash-dotted line is the resonance wavelength shift from the higher order mode. The fundamental WGM at 1549.6 nm is shifted as much as 0.4 nm, while the second order WGM at 1569 nm experiences negligible shift.

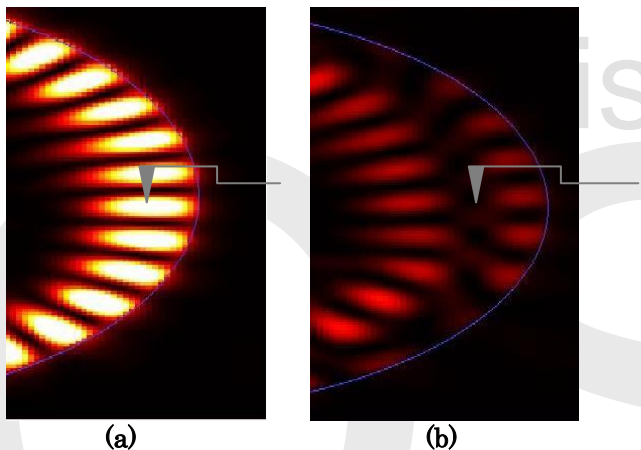

Fig. 3: The mode profile (electromagnetic field intensity) of the fundamental optical resonance mode (a) and the 2nd order optical resonance mode (b), excited by the coupling bus waveguide

The mode profiles were simulated using the Finite Element Method. Fig. 3 illustrate the mode profile of the fundamental mode and the 2nd order mode, it clearly shows the drastic different mode characteristics of these two excited modes. The different mode profile also explains our observations why the resonance wavelength shifts of the two modes are so different when the AFM tip is placed at the nodal point of the 2nd order resonance mode. At the position where the fundamental mode has close to maximum electromagnetic energy, because there is almost no electromagnetic energy (node) at the position for the higher order mode, placing an AFM tip close to the point causes the relevant large resonance wavelength shift of the fundamental mode and almost no shift for the 2nd order mode. The higher order mode nodal detection is potential for making ultra sensitive integrated micro- and nano-scale devices with many applications in biomedicine, food and security areas. As nanotechnology has enabled us to make various nano scale objects with the dimension from a few nanometers to several hundred nanometers, how to make extremely sensitive and small size detection devices is critical for us to have a better understanding and further control the nano scale objects. The self referencing detection mechanism from the different modal characteristics of the fundamental mode and the higher order mode provides us a detection mechanism independent of the external perturbation. In future works, it will be interesting to show that a temperature change will influence the resonances for both the fundamental and the 2nd-order modes, while a nanoparticle in the right vicinity only influences one but not the other.

In summary, we have demonstrated the detection of the higher order optical resonance node using a integrated *disk* microresonator. There is a negligible shift for the 2nd order mode, compared to the nearly maximum shift induced by the fundamental mode, excited simultaneously by the coupling bus waveguide. With such different optical resonance mode behavior, one potentially has the capability to make a self referencing optical detection and sensing devices.